\documentclass[12pt] {article} \usepackage{graphicx} \usepackage{epsfig}
\begin {document} \parindent=15pt
\begin{center} {\bf HEAVY FLAVOUR PRODUCTION IN $pp$ AND HEAVY ION COLLISIONS
IN QCD UP TO LHC ENERGIES}\\ \vspace{.5cm} C. Merino$^*$, C. Pajares$^*$,
M.M. Ryzhinskiy$^{**}$, Yu.M. Shabelski$^{**}$, 
and A.G. Shuvaev$^{**}$ \\
\vspace{.5cm} $^*$ Departamento de F\'\i sica de Part\'\i culas, Facultade
de F\'\i sica \\ and Instituto Galego de F\'\i sica de Altas Enerx\'\i as (IGFAE)\\
Universidade de Santiago de Compostela\\
15782 Santiago de Compostela, Galiza, Spain \\
\vspace{.2cm}
$^{**}$ Petersburg Nuclear Physics Institute \\
Gatchina, St.Petersburg 188350, Russia \\
\end{center}
\vspace{.5cm}

\begin{abstract}

Charm production in $pp$ collisions is considered in the framework of perturbative 
QCD. The values of two parameters, the charm quark mass $m_c$ and the QCD scale $\mu^2$,
are determined from the comparison of the theoretical calculations with experimental data.
The RHIC data on charm and beauty production are compared with the $k_T$-factorization approach  
predictions and with standard NLO QCD. The calculated results underestimate the 
STAR Collaboration data. The role of possible nuclear effects is discussed.  
The predictions for LHC energies are also given.

\end{abstract}


\newpage

\section{Introduction}

The investigation of heavy quark production in high energy collisions is an
important method for studing the quark-gluon structure of hadrons and the
possible nuclear effects at early stages of secondary production. The 
description of hard interactions in hadron collisions within the framework of 
QCD is possible only with the help of some phenomenological assumptions which 
reduce the hadron--hadron interaction to the parton--parton one via the formalism of 
the hadron structure functions. The cross sections of hard processes in hadron--hadron 
interactions can be written in a factorized form \cite{CSS} as the convolutions of 
the squared matrix elements of the subprocess calculated within the framework of 
QCD with the parton distributions in the colliding hadrons.

The most popular phenomenological approach is the NLO QCD collinear approximation 
[2--5], where the cross sections of QCD subprocesses are calculated in the 
Next-to-Leading Order (NLO) of $\alpha_s$ series. The Fixed Order plus 
Next-to-Leading-Log (FONLL)~\cite{CGN} also resumes large perturbative terms 
proportional to $\alpha_s^n\cdot \log^k(p_T/m)$ with $k = n, n-1$, where $p_T$ and $m$
are the heavy quark transverse momentum and the mass, respectively. In these calculations
all particles involved are assumed to be on mass shell, carrying only longitudinal momenta, and 
the cross section is averaged over two transverse polarizations of the incident 
gluons. The virtualities $q^2$ of the initial partons are taken into account only through
their structure functions.

The formalism which incorporates the incident parton transverse momenta is 
refered to as the $k_T$-factorization approach [7--10], or the theory of 
semihard interactions [11--19]. Here the Feynman diagrams are calculated 
taking into account the virtualities and all possible polarizations of the 
incident partons. In the small $x$ domain there is not ground to neglect the 
transverse momenta of gluons, $q_{1T}$ and $q_{2T}$, when compared to the quark mass 
and to the quark transverse momenta. In such an approach the QCD matrix elements
of the partonic subprocesses are rather complicated. We have calculated them 
in LO. On the other hand, the multiple emission of soft gluons has been included
in our calculation. In the $k_T$-factorization approach the unintegrated gluon
distributions are used instead of the usual structure functions.

For the case of heavy ion collisions the linear A-dependence of the heavy flavour 
production cross sections is usually assumed, so the obtained results are ussually
discussed in terms of ``cross section scaled to $pp$ collision". This general assumption
is experimentally supported by the result of \cite{Lei}, $\sigma \sim A^{\alpha}$
with $\alpha = 1.02 \pm 0.03 \pm 0.03$ (recent results of SELEX Collaboration
\cite{SELEX} give significantly smaller values of $\alpha$). In perturbative QCD with
the factorization approximation \cite{CSS} the case of heavy ion collisions differs from
$pp$ collisions only by the change of the usual nucleon structure functions or unintegrated
gluon distributions by the same functions for bound nucleons. At RHIC energies the difference
among parton structure functions is known from EMC and NMC experimental data \cite{Arn} and their
GLDAP evolution equation analysis \cite{Esk}. As a result, the nuclear effects in QCD for the
total cross section of charm production at RHIC are estimated to be on the level of 5--10 \% \cite{APSS}
of the linear A-dependence values. 

The experimental data of PHENIX \cite{PHE,PHE1} Collaboration obtained at RHIC 
both from $pp$ and nuclear collisions are in reasonable agreement with the NLO QCD
and the $k_T$-factorization approach predictions. At the same time, the STAR \cite{STAR}
Collaboration data obtained at RHIC in nuclear collisions are in contradiction with
standard QCD calculations. In the present paper we discuss the accuracy of the theoretical
calculations, and the plausible reasons for the charm production enhancement on nuclear targets.

\section{Determination of the QCD parameters}

The standard QCD expression for the heavy quark production cross section in a 
hadron~1--hadron~2 collision has the form 
\begin{equation}
\sigma^{12\rightarrow Q\overline{Q}} = \int_{x_{a0}}^{1} dx_a
\int_{x_{b0}}^{1} dx_b G_{a/1}(x_a,\mu^2_F) \cdot G_{b/2}(x_b,\mu^2_F) 
\cdot \hat{\sigma}^{ab\rightarrow Q\overline{Q}}(\hat{s},m_Q,\mu^{2}_R) \;,
\end{equation}
where $\mu_F$ is the QCD factorization scale, 
$x_{a0} = \textstyle 4m_Q^2/ \textstyle s$, and 
$x_{b0} = \textstyle 4m_Q^2/ \textstyle (sx_a)$. Here $G_{a/1}(x_a,\mu^2_F)$ 
and $G_{b/2}(x_b,\mu^2_F)$ are the structure functions of partons $a$ and $b$ 
inside hadrons $1$ and $2$, respectively, and 
\begin{equation}
\hat{\sigma}^{ab\rightarrow Q\overline{Q}}(\hat{s},m_Q,\mu^{2}_R) =
\alpha^2_s(\mu^2_R)\cdot\sigma_{ab}^{(0)} + \alpha^3_s(\mu^2_R)\cdot\sigma_{ab}^{(1)}
\end{equation}
is the cross section of the subprocess $ab\rightarrow Q\overline{Q}$, given 
by standard QCD as the sum of LO and NLO contributions \cite{NDE,Alt,Bee}. 
These contributions depend on the parton center-of-mass energy 
\mbox{$\hat{s} = (p_a+p_b)^2 = x_a\cdot x_b\cdot s$}, the mass of the produced
heavy quark $m_Q$ (actually only on $\rho = 4m^{2}_{Q}/\hat{s}$), and the QCD
renormalization scale $\mu^2_R$.  

Thus in the QCD calculation of total cross section of charm production three
parameters are involved: the heavy quark mass, $m_Q$, and the two QCD scales,
$\mu^2_F$ and $\mu^2_R$. The heavy quark mass $m_Q$ can be estimated as a half
of the mass of light quarkonium. More accurately, in QCD the masses of heavy
quarks should depend on the distance at which they are measured. When including
this dependence one obtains \cite{Nar,BBB}:
\begin{equation}
m_c = 1.4  \; {\rm GeV} \;, \;\; m_b = 4.6 \; {\rm GeV} \;.
\end{equation}

The two QCD scales are usually assumed to be equal in numerical calculations, 
$\mu^2 = \mu^2_R = \mu^2_R$, and from a general point of view they should be of the 
order of the maximal hardness of the interaction. If the produced heavy quarks are
heavy enough, one can assume that
\begin{equation}
\mu^2 \simeq m^2_Q \;,
\end{equation}
though the estimation $\mu^2 \simeq \sqrt{m^2_Q + p^2_T}$ used in the 
$k_T$-factorization approach seems to be also reasonable.

The next question concerns the uncertainties in the values of these parameters.
Usually, the accuracy of our knowledge on heavy quark mass is assumed
to be about 10\% \cite{FMNR} (for $c$-quark), and smaller for $b$-quark.
The general thinking is that the uncertainty of the value of
QCD scale $\mu^2 \simeq m^2_Q$ is about a factor 2 larger.
Of course, there is no reason why factor 2 should be more reasonable than, 
say, a factor 1.5 or a factor 3. For our purpose, the uncertainties in the values
of the discussed parameters should be in any case compatible with the values of
experimental data including their errorbars. As an example, the low boundary for the
charm production cross section presented in \cite{Ramo} is in total disagreement
with all available experimental data, meaning that the uncertainties assumed for the
values of the parameters were too large.

We obtain the values of the QCD parameters $m_c$ and $\mu^2$ by fitting the
results of the NLO QCD expression in Eqs.~(1) and (2) to the 
fixed target experimental data in \cite{Lou} and \cite{Abr}, the RHIC point 
for $pp$ collisions \cite{Mori}, and the UA2 Collaboration point \cite{UA2}.
We have used the GRV95 parton distributions \cite{GRV}, which are compatible
with the most modern analysis (see discussion in \cite{GRV1}). Moreover, the difference
to calculations performed with other sets of parton structure functions is small 
in the considered energy region. The result of our fit is shown in Fig.~1 by solid curve. 
\begin{figure}[htb]
\centering
\vskip .5cm
\includegraphics[width=.65\hsize]{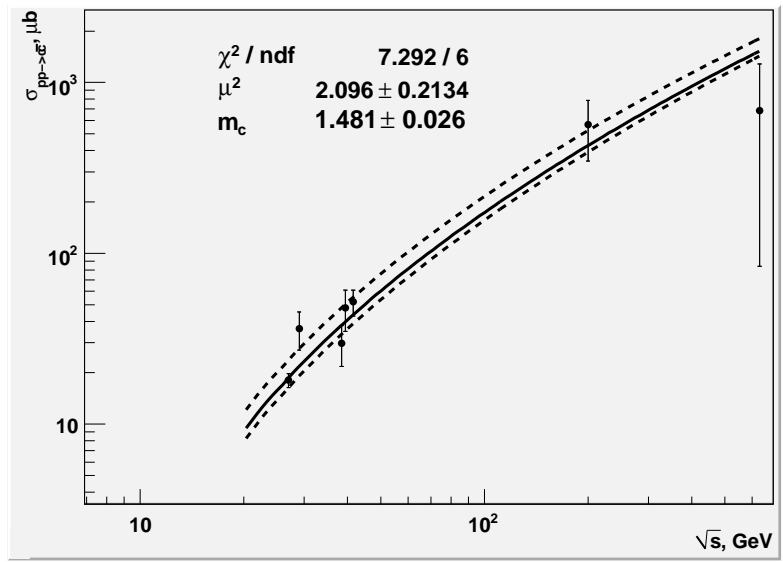}
\caption{
Total cross section of charm production in $pp$ and $\bar{p}p$ collisions at 
$\sqrt{s} > 25$~GeV. Experimental fixed target data are taken from \cite{Lou} and 
\cite{Abr}, RHIC point from \cite{Mori}, and UA2 Collaboration data from \cite{UA2}.
The solid curve shows the result of the fit of QCD parameters $m_Q$ and $\mu^2$,
and dashed curves show the edges of the uncertainties
(the region between the $\chi^2$ minimal value and the $\chi^2$ minimal value
plus one).}
\end{figure}

This fit corresponds to the following parameter values:
\begin{equation}
m_c = 1.48 \pm 0.03 \; {\rm GeV} \;, \;\; \mu^2 = 2.10 \pm 0.21 \; {\rm GeV}^2 \;,
\end{equation}
that seem to be very natural. The value of $\chi^2$ is 7.3/6 ndf. The uncertainties are 
shown by dashed curves and they correspond to increase the value of $\chi^2$ by one from 
its minimal value.

The same analysis for the total cross section of charm production in $\pi p$
collisions is shown in Fig.~2. Here the value of $\chi^2$ is 12./7 ndf and the 
parameter values are
\begin{equation}
m_c = 1.83 \pm 0.30 \; {\rm GeV} \;, \;\; \mu^2 = 2.90 \pm 0.06 \; {\rm GeV}^2 \;,
\end{equation}

One has to note that in both analysis we obtain $\mu^2 \simeq m^2_c$, as it is usually
assumed {\it a priori}. Also the two values of $m_c$ almost coincide inside the error bars.
On the other hand, the difference seeming to appear in the $\mu^2$ values 
can be connected to the not very good knowledge of the pion structure functions.
\begin{figure}[htb]
\centering
\vskip .5cm
\includegraphics[width=.55\hsize]{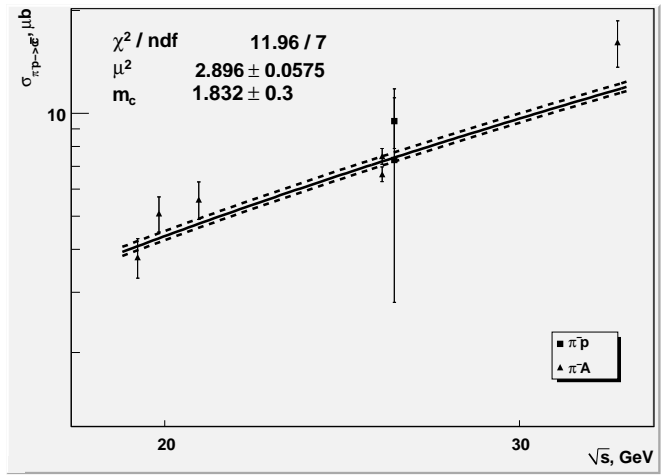}
\caption{Total cross section of charm production in $\pi^{\pm}p$ collisions at 
$\sqrt{s} > 18$~GeV. Experimental fixed target data are  taken from \cite{Lou} 
and \cite{Abr} (they are all shown by points). The solid curve shows the result
of the fit of QCD $m_Q$ and $\mu^2$, and dashed curves show the edges of the uncertainties
(the region between the $\chi^2$ minimal value and the $\chi^2$ minimal value
plus one).}
\end{figure}
\vspace{-0.5cm}

\section{Charm production in different QCD approaches}

The experimental data on the total cross section of charm production at high energies
shown in Fig.~1 are presented again in Fig.~3, together with data on
interactions with nuclei in cosmic rays~\cite{Xu} and RHIC nuclear 
data~\cite{PHE,PHE1,STAR} scaled to $pp$ interactions. The last cross section 
data obtained by the STAR Collaboration, as well as cosmic rays date, are larger than the 
theoretical NLO QCD predictions shown by the dash-dotted curve. One can also see the 
different data obtained by the PHENIX and the STAR collaborations. This discrepancy is 
discussed in detail in references~\cite{Xu,Sua}.
\begin{figure}[htb]
\centering
\includegraphics[width=.55\hsize]{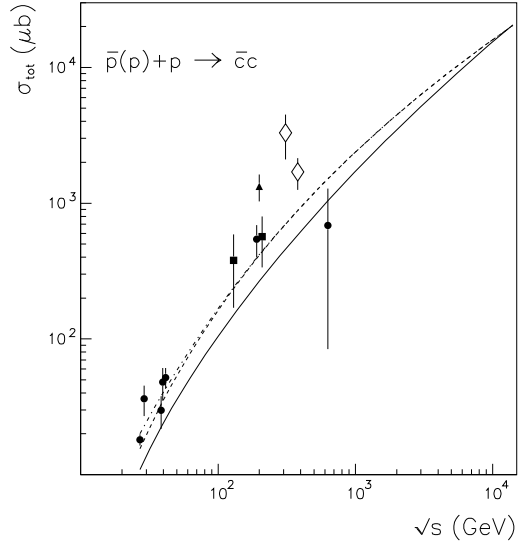}
\vskip -.5cm
\caption{
Total cross section of charm production in $pp$ and $\bar{p}p$ collisions at 
$\sqrt{s} > 25$~GeV. Experimental fixed target data are  taken from \cite{Lou} 
and \cite{Abr}, the RHIC point \cite{Mori}, and the UA2 Collaboration data from \cite{UA2} 
(all are shown by points). PHENIX data \cite{PHE,PHE1} and STAR data \cite{STAR} in nuclear collisions
are shown by squares and triangles, respectively, while cosmic rays data \cite{Xu} are shown by diamonds.
The solid curve shows the theoretical result obtained in the $k_T$-factorization approach, the
dashed curve corresponds to NLO QCD with only gluon-gluon fusion contribution, and the dash-dotted curve
corresponds to the total NLO QCD result.}
\end{figure}

All other experimental points are in reasonable agreement with the NLO QCD 
calculations performed by taking the values of the parameters presented in Eq.~(5) (dash-dotted curve),
and where the GRV95 parton distributions, compatible with the most modern analysis
(see discussion in \cite{GRV1}), were used. 

At high energies, the main contribution to the heavy flavour 
production cross section comes from the small $x$ region. Here there is not ground to neglect
the transverse momenta of gluons, $q_{1T}$ and $q_{2T}$, with respect to the quark mass and to the quark
transverse momenta, $p_{iT}$. In particular, at high energies and high $p_{iT}$ the main contribution to the cross
sections comes from the region of $q_{1T} \sim p_{1T}$, or of $q_{2T} \sim p_{1T}$ (see \cite{RSS} for details).
 
For our calculations in the framework of the $k_T$-factorization approach we have considred the
heavy quark $Q=c, b$ mass \cite{SS} as $m_c =$ 1.4 GeV and $m_b =$ 4.6 GeV, and we have used
QCD scales $\mu^2_R = \mu^2_F = m^2_T$, $m^2_T = m^2_Q + p^2_T$. All details of the calculations
in the $k_T$-factorization approach presented below can be found in \cite{SS}. The result of these
calculations are shown by a solid curve in Fig.~3, where one can see that the $k_T$-factorization approach
gives at not very high energies cross sections which are smaller than those obtained by NLO QCD calculations,
but which, on the other hand, increase slightly faster with energy.  

The cross section of beauty production in $pp$ collisions at RHIC energy
$\sqrt{s} = 200$ GeV has been measured by PHENIX Collaboration \cite{09034851}
to be $\sigma_{b\bar{b}} = 3.2^{+1.2}_{-1.1}$ $^{+1.4}_{-1.3}$ $\mu$b, what is in agreement with
the result of our NLO QCD calculation $\sigma_{b\bar{b}} = 3.06 \mu$b, obtained at the same energy 
$\sqrt{s} = 200$ GeV by using GRV95 parton distributions.

The cross sections of heavy flavour production at RHIC are obtained by the 
extrapolation of the data measured in the central region to the whole region of 
rapidities. So it seems reasonable to compare the theoretical results to the experimental data 
before extrapolation, what will allow, as a minimum, to avoid a part of the uncertainties.
For this sake, we present in Tables 1 and 2 the experimental values of 
$d\sigma/dy$ ($\vert y \vert = 0.$) for charm and beauty production, together with
the theoretical values obtained both in NLO QCD calculations with GRV95 parton
distributions and in the $k_T$-factorization approach. Again, the STAR data for charm
production appear to be several times larger than the theoretical results, whereas the PHENIX
data for both charm and beauty production are in a reasonable agreement with theory.
\begin{center}
\begin{tabular}{|c||r|r|} \hline
Collaboration/ & Reaction\hspace{1.cm} &  $d\sigma/dy$ ($\vert y \vert = 0)$  \\ 
Theory         &                   &    \\  \hline
PHENIX         & Au--Au (m.b.) \cite{0409028} & $143 \pm 13 \pm 36 \mu$b   \\ 
               & p--p  \cite{PHE1}\hspace{1.cm} & $123  \pm 12 \pm 45 \mu$b  \\  \hline
STAR           & d--Au \cite{0607011}\hspace{0.7cm} & $301 \pm 44 \pm 67 \mu$b   \\
               & Au--Au (m.b.) \cite{0607011} & $267 \pm 19 \pm 49 \mu$b   \\
             & Au--Au (central) \cite{0607011}  &  $283 \pm 12 \pm 39 \mu$b  \\
             & Cu--Cu (m.b.) \cite{0709.4223}  &   $260 \pm 34 \mu$b\hspace{0.3cm}  \\ \hline
NLO QCD      & p--p\hspace{1.5cm} & 83$\mu$b\hspace{1.cm} \\ \hline
$k_T$-factorization & p--p\hspace{1.5cm} & 56$\mu$b\hspace{1.cm} \\ \hline
\end{tabular}
\end{center}
Table 1. The experimental and theoretical values of $d\sigma/dy$ ($\vert y \vert = 0.$)
for charm production at the RHIC energy $\sqrt{s} = 200$ GeV, scaled to $pp$.

\newpage

\begin{center}
\begin{tabular}{|c||r|r|} \hline
Collaboration/ & Reaction &  $d\sigma/dy$ ($\vert y \vert = 0)$  \\ 
Theory         &                   &    \\  \hline
PHENIX         & p--p \cite{09034851} 
& $0.92^{+0.34}_{-0.31}$ $^{0.39}_{-36} \mu$b   \\ 
\hline
NLO QCD      & p--p\hspace{0.6cm} & 0.97$\mu$b\hspace{1.cm} \\ \hline
$k_T$-factorization & p--p\hspace{0.6cm} & 1.62$\mu$b\hspace{1.cm} \\ \hline
\end{tabular}
\end{center}
Table 2. The experimental and theoretical values of $d\sigma/dy$ ($\vert y \vert = 0.$)
for beauty production at the RHIC energy $\sqrt{s} = 200$ GeV, scaled to $pp$.
\vskip 0.4cm

In Table~3 we present the predictions for heavy flavour production cross sections and
midrapidity inclusive densities that we have obtained in the framework of $k_T$-factorization
approach.
\begin{center}
\begin{tabular}{|c||r|r|r|r|} \hline
Reaction & Quantity\hspace{0.5cm} & $\sqrt{s}$ = 5.5 TeV & $\sqrt{s}$ = 7 TeV & $\sqrt{s}$ = 14 TeV  \\  \hline
& & & &\\
$pp \to c\bar{c}$ & $\sigma$\hspace{1.2cm} & 9 mb\hspace{0.7cm} & 11.3 mb\hspace{0.5cm} & 21 mb\hspace{0.6cm} \\
\cline{2-5}
& $d\sigma/dy$ ($\vert y \vert = 0)$ & 1.1 mb\hspace{0.5cm} & 1.35 mb\hspace{0.5cm} & 2.3 mb\hspace{0.5cm} \\
& & & &\\
\hline
& & & &\\
$pp \to b\bar{b}$ & $\sigma$\hspace{1.2cm} & 0.44 mb\hspace{0.4cm} & 0.57 mb\hspace{0.4cm} & 1.2 mb\hspace{0.5cm} \\
\cline{2-5}
& $d\sigma/dy$ ($\vert y \vert = 0)$ & 0.063 mb\hspace{0.3cm} & 0.079 mb\hspace{0.3cm} & 1.5 mb\hspace{0.5cm} \\
& & & &\\
\hline
\end{tabular}
\end{center}
Table 3. The predictions for charm and beauty production cross sections 
$\sigma$, and for the inclusive densities $d\sigma/dy$ ($\vert y \vert = 0.$), 
obtained in the $k_T$-factorization approach for $pp$ collisions at LHC energies.

\section{Transverse momentum distributions} 

The STAR Collaboration also presents the $p_T$-distributions of $D$-mesons 
produced in d+Au collisions and scaled to $pp$ collisions at $\sqrt{s}$ = 
200 GeV. These data, taken from \cite{CNV}, are presented in Fig.~4 together
with the corresponding theoretical calculations. 
\begin{figure}[htb]
\centering
\includegraphics[width=.55\hsize]{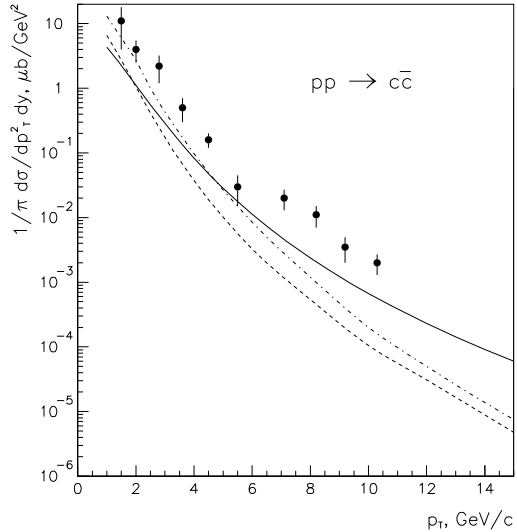}
\vskip -.5cm
\caption{
The STAR Collaboration data for $p_T$-distributions of $D$-mesons produced in
d+Au collisions and scaled to $pp$ collisions at $\sqrt{s}$ = 200 GeV,
together with the corresponding calculations in the $k_T$-factorization approach
(solid curve), NLO QCD (dashed curve), and FONLL \cite{CNV} (dash-dotted curve).}
\end{figure}

The NLO QCD result shown by the dashed curve is in an evident disagreement (significantly below)
with the experimental data on the same level as in Fig.~3. The upper curve of FONLL calculations
for charmed quark production taken from \cite{CNV} (dash-dotted curve) also underestimates
the data. The result of the $k_T$-factorization approach (solid curve) presents a reasonable 
slope in $p_T$-dependence, but also it underestimates the data. In this case, the level 
of disagreement at high $p_T$ is, however, smaller than in the case of the 
total cross sections, what should be
connected with the fact that a rather large contribution to the total cross 
section comes
from the low-$p_T$ region ($p_T \leq m_c$), where both experimental and theoretical
uncertainties are rather large.

It is necessary to note that in Fig.~4 we compare the experimental points for 
$D$-meson production to theoretical curves for charmed quark distributions. 
In contradiction to what is claimed in \cite{CNV}, on top of fragmentation
processes where the momentum of the $D$-meson is smaller than the momentum
of the $c$-quark, also recombination processes where the momentum of $D$-meson
is larger than the momentum of $c$-quark are active. The existence of recombination
processes in charm production seems to be evident from the experimental data on the 
asymmetry in yields of the so-called favoured and unfavoured $D$-mesons (see 
discussions in [45--48]). As a matter of fact, though the produced heavy and light 
quarks have very different transverse momenta, the difference in the 
components of their velocities can be not so large. It is very possible that the 
fragmentation and recombination processes in charm quark hadronization balance 
each other in the processes with not very high $p_T$, and as a hint of this,
the calculated Feynman-$x$ distributions of produced charm quarks in $\pi p$
collisions are in good agreement with the experimental distributions of produced
$D$-mesons (see Fig.~5 in \cite{FMNR}).

The total cross section of charm production was not measured at 
Tevatron-collider energies. However, data on $p_T$-distributions of 
$D$-mesons exist at these energies \cite{Aco}. They are presented in
Fig.~5, where one can see that they are in good agreement with NLO QCD
calculations that use GRV95 parton distributions for charm quarks (dashed curve). 
\begin{figure}[htb]
\centering
\includegraphics[width=.55\hsize]{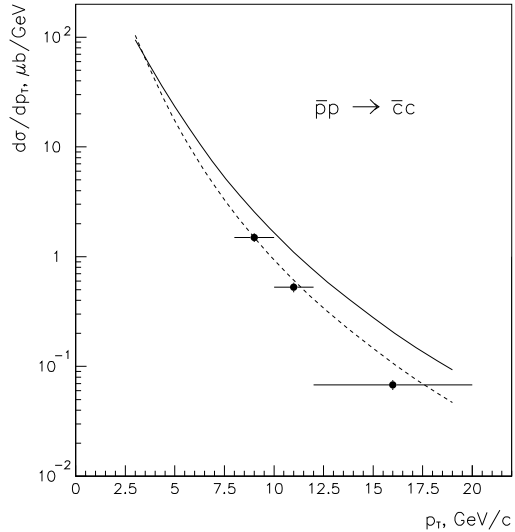}
\vskip -.5cm
\caption{
The cross sections for $D$-meson production in $\bar{p}p$ collisions at 
$\sqrt{s}$ = 1.96 TeV~\cite{Aco} with $\vert y_1 \vert \leq 1$, together with 
calculations in the $k_T$-factorization approach (solid curve) and in NLO QCD
(dashed curve).}
\end{figure}

The solid curve corresponds to the $k_T$-factorization approach and it slightly
overestimates the data, but the agreement should become better in the future when 
the contribution of charmed antibaryons will be added to the experimental data. 
The results of FONLL calculations \cite{CNV} with fragmentation functions for 
$D$-mesons production presented in \cite{Aco}, slightly underestimate the yields
of $D$-mesons. Thus, all QCD approaches (see also \cite{Ram}) are in reasonable 
agreement with the experimental data at $\sqrt{s}$ = 1.96 TeV. Some disagreement 
of these data with \cite{DY} can be expected by the QCD parameter values used there,
where the charm production cross section increases with energy very fast. This
should probably result in a rather good agreement of \cite{DY} with the STAR data
presented in Fig.~4, but also in the overestimation of the Tevatron-collider data
presented in Fig.~5. 

Finally, the theoretical predictions for $d\sigma/dp_T$ distributions of charm and beauty
quarks produced in the rapidity window $\vert y_1 \vert \leq 1$ at three LHC energies,
$\sqrt{s}$ = 5.5 TeV, 7 TeV, and 14 Tev, calculated in the $k_T$-factorization approach
are presented in Fig.~6.
\begin{figure}[htb]
\centering
\includegraphics[width=.55\hsize]{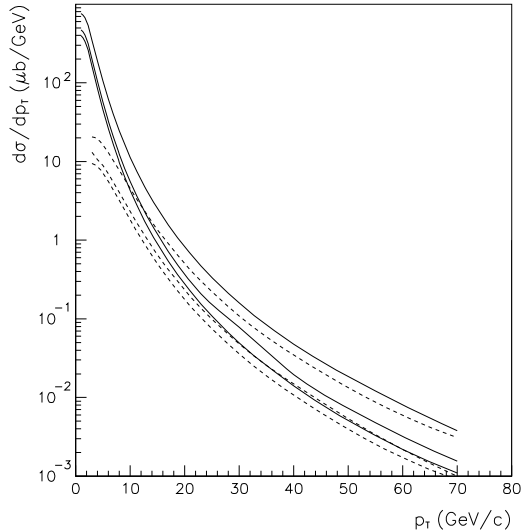}
\vskip -.5cm
\caption{
The predicted cross sections of charm (solid curves) and beauty (dashed curves) production
in $pp$ collisions at LHC energies, from bottom to top, $\sqrt{s}$ = 5.5 TeV, 7 TeV, and 14 Tev,
in $\vert y_1 \vert \leq 1$, obtained in the $k_T$-factorization approach.}
\end{figure}

\section{Nuclear effects in charm production at RHIC}
 
First, it can be said that the EMC-effect, i.e. the nuclear deformation of 
parton distributions \cite{Esk}, should increase the total cross sections for 
charm production calculated in the NLO QCD linear approximation at RHIC 
energy by 5--10 \% \cite{APSS} with respect to the linear A-dependence.

The data of PHENIX and STAR Collaborations scaled to binary interactions are 
presented together in \cite{Sua}. There one can see the absence of nuclear effects
in the PHENIX data from $pp$ to central Au-Au collisions, and in STAR data from
d-Au to central Au-Au collisions, in both cases at the level of 20\% accuracy.
Concerning the PHENIX data presented in \cite{Xu}, they do not present any visible
dependence of the charm production cross section on the number of collision
participant nucleons, $N_{part}$, in Au-Au collisions at different centralities.

One factor $\sim$2--3 discrepancy \cite {Sua} exists between the total cross 
sections of charm production obtained by PHENIX and STAR Collaborations.
It seems evident that the explanation of this discrepancy could be completely
connected to some experimental problem. However, meanwhile the experimental
collaborations will not clarify this point, two scenarios can be scripted:
either the PHENIX data are right and the NLO QCD calculation reasonably describes
all experimental data except for two cosmic ray points presented in Fig.~3,
and then nuclear effects in total cross sections of charm production are small,
or, on the contrary, the STAR data are correct and the nuclear effects increase
abot 4--5 times the total cross sections of charm production. In the case the
$N_{part}$ dependence of these nuclear effects would saturate very fast, even the
cosmic ray data could be included in the theoretical description.

In this second scenario, so large nuclear effects could be connected to large
non-perturbative contributions (which could be larger than perturbative contributions)
to high density states, e.g. string fusion \cite{MPR}, percolation \cite{BP}, strong
colour field \cite{Topor}, or colour
glass condensate effects \cite{GV,KT} in the interactions with nuclei. 

In these non-perturbative approaches, the strong colour field inside the cluster formed
by the overlapping strings produces, above some scale $\eta_c$ \cite{BP} given by the
critical percolation string density, heavy $Q\bar{Q}$ pairs via the Schwinger 
mechanism in the same way as a single string produces light $q\bar{q}$ pairs. Thus, 
in the colour glass condensate approach \cite{GV,KT} the significant scale is 
the saturation momentum $Q_s$, which grows with energy and nuclear size. When
$Q_s > m_T(c\bar{c})$, the classical colour field is strong enough and it produces
heavy pairs $Q\bar{Q}$. In summary, the production pattern for heavy quarks in these
non-perturbative approaches becomes similar to that of the light quarks and so an
overall enhancement of heavy quark production cross section is expected.
However, to get numerical estimation of these nuclear effects a set of additional
assumptions is required.

\vspace{-0.2cm}
\section{Conclusion}

We show that the $k_T$-factorization approach predictions are in reasonable agreement
with the NLO QCD calculations for the total cross section of charm and beauty production
at RHIC energies.

We obtain a reasonable description of the PHENIX Collaboration data including the points
obtained from nuclear collisions, and our results are clearly in contradiction with
the STAR Collaboration data.

It seems that the main disagreement of our calculations with the STAR Collaboration data
comes from the low-$p_T$ region ($p_T \leq 1$ GeV/c), where both experimental and
theoretical uncertainties are rather large.

The predicted $p_T$-distribution of the produced charm at high $p_T$ in the
$k_T$-factorization approach is higher than the NLO QCD and FONLL predictions.
The $k_T$-factorization approach only slightly underestimates the experimental
data of the STAR Collaboration.

Since the NLO QCD calculation basically agrees with the PHENIX data, one
would be in principle prompted to think that the PHENIX data are right. Either way,
the question of the existence of a factor $\sim$2--3 discrepancy between the total
cross section of charm production obtained by PHENIX and by STAR Collaborations has
to be experimentally unraveled in order to understand the basic mechanisms working
in charm production.

The predictions for LHC energies are also presented.
\vskip 0.2cm

{\bf Acknowledgements}

We are grateful to N. Armesto, A. Khodjamirian, and M. G. Ryskin
for useful discussions. This work was supported by Ministerio Educaci\'on y
Ciencia of Spain under project FPA 2005--01963, and by Xunta de Galicia. It
was also supported by grant RSGSS--3628.2008.2 

\end{document}